# Crystal growth of $MgB_2$ from Mg-Cu-B melt flux and superconducting properties

D. Souptel[1, *], G. Behr[1], W. Löser[1], W. Kopylov[2] and M. Zinkevich[1, I)]

[1]Leibniz-Institut für Festkörper- und Werkstoffforschung Dresden, Postfach 270016, D-01171 Dresden, Germany

[2]Institute of Solid State Physics RAS, 142432 Moscow District, Chernogolovka, Russia

## Abstract

A new method for preparation of single crystals of the superconducting intermetallic $MgB_2$ compound from a Mg-Cu-B melt flux is presented. The high vapour pressure of Mg at elevated temperature is a serious challenge of the preparation process. The approximate thermodynamic calculations of the ternary Mg-Cu-B phase diagram show a beneficial effect of Cu, which extends the range of formation of $MgB_2$ to lower temperatures. Within the as-solidified Mg-Cu-B melt flux the $MgB_2$ compound forms plate-like single crystals up to a size of $0.2 \times 0.2 \times 0.05$ $mm^3$ or alternatively rims peritectically grown around $MgB_4$ particles. AC-susceptibility measurements were conducted with specimen selected from different parts of the as-solidified flux containing $MgB_2$ particles. Peritectically formed $MgB_2$-particles display the highest transition temperature of $T_c$ = 39.2 K and a relatively narrow transition width of $\Delta T_c$ = 1.3 K. Other sections of the sample exhibit various superconducting transitions from $T_c$ = 39 K to 7.2 K. This variation of $T_c$ is attributed to a finite homogeneity range of the $MgB_2$ compound whereas significant Cu solid solubility in $MgB_2$ can be excluded.

[I)] Present address: Max-Planck-Institut für Metallforschung, Heisenbergstr. 3, D-70569; Stuttgart, Germany

## 1. Introduction

The recent discovery of superconductivity of the intermetallic compound $MgB_2$ [1] has aroused a flurry of theoretical and experimental work [2, 3, 4, 5, 6]. With $T_c$ = 39 K, $MgB_2$ exhibits the highest superconducting transition temperature of intermetallic compounds, which is nearly twice as high as the top $T_c$ values of A15 compounds ($Nb_3Ge$: $T_c$ =23.2K) and Rare Earth-Transition Metal-Borocarbides ($YPd_2B_2C$: $T_c$ = 23 K). Only the ceramic oxide superconductors and compounds based on $C_{60}$ exceed these $T_c$ value [7].

Many important parameters of $MgB_2$ have already been derived, such as the upper critical field ($H_{c2}$ =13 – 20.4 T) [2, 3], the Ginzburg-Landau parameter (k ≈ 26) [3] and the bulk critical superconducting current density ($j_c = 8 \cdot 10^4$ $Acm^{-2}$ at 4.2 K and 12 T) [4] in thin films. The mechanism of superconductivity in $MgB_2$ is still an open question. Many experiments support the view that $MgB_2$ is a phonon-mediated superconductor in the weak- to moderate-coupling regime [5]. On the other hand it is proposed that the theory of hole superconductivity may also account for $MgB_2$ [6]. Experimental data that can shed light on its synthesis, doping, phase formation and the phase equilibria in this system are of great interest. One of the surprising features of superconducting $MgB_2$ is the simplicity of both its chemistry and structure. $MgB_2$ crystallizes in the $AlB_2$ crystal structure where honeycomb layers of boron atoms alternate with hexagonal layers of Mg atoms. The lattice parameters are a = 0.30834 nm and c = 0.35213 nm [8]. But the very limited substitutional chemistry [9,10,11] possibly suggests that the structure may be more complex than on a first glance.

Although this material is available from common chemical suppliers and has been characterized since the mid 1950s, up to now the investigations are basically limited to polycrystalline samples and thin films [4]. Ceramic samples were successfully prepared by ambient pressure synthesis in sealed Ta or Mo crucibles [3,12], but more dense ceramic was obtained by hot isostatic pressing at 950°C under pressure of several GPa of boron and magnesium powders as starting materials [1,13]. There were great efforts in doping of $MgB_2$ [14,15]. There is also an idea of a finite homo-geneity range of $MgB_2$ compound. A higher $T_c$ is expected if the composition is on the B-rich side [16], which is in accord with the very low $T_c$ = 12 K of Mg-rich films [17]. Only recently growth of $MgB_2$ single crystals with maximum size 0.5 × 0.5 × 0.02 $mm^3$ has been achieved using a vapour transport method [18]. The crystals show a relatively gradual superconducting transition with an onset temperature of 38.6 K. Growth from the melt, which can in principle lead to larger sized crystals, is aggravated by the high vapour pressure of Mg. To our knowledge, there was no successful attempt of $MgB_2$ crystal growth from Mg-B melts so far.

In this work, a new method of $MgB_2$ crystal growth from the Mg-Cu-B flux is presented. The reasons for the choice of Cu as a solvent for B and Mg are: (i) large solubility of boron in the Cu-B melt at least 13.3 at.% at the eutectic temperature 1013°C, (ii) absence of binary Cu-B [19] or ternary Mg-Cu-B intermetallic compounds and improbable doping of $MgB_2$ with Cu because of the high affinity of Cu and Mg to form intermetallic compounds [14] (iii) facilitated crystallization of desired refractory magnesium borides in Mg-Cu-B because of low melting temperatures (< 800°C) of remaining binary compounds. Our experimental trials are supported by approximate thermodynamic calculations of the ternary Mg-Cu-B phase diagram, which are based on existing thermodynamic data of binary systems. The phase diagram data served as a guideline for optimizing process parameters of crystallization and interpretation of various microstructures observed in Mg-Cu-B fluxes in a wide composition range. The results of measurements of superconducting properties of Mg-Cu-B samples containing $MgB_2$ crystals with different morphologies are presented.

## 2. Experimental methods

### *2.1. Crystal growth procedure*

The Mg-Cu-B flux was prepared from Cu (99.99%), Mg (99.9%) powders (or cheeps) and commercial $MgB_2$ powder (Alfa Aesar) as a source of boron. Using $MgB_2$ powder is superior with respect to pure amorphous boron due to its higher reactivity with the melt, smaller grain size and higher purity (especially: the low oxygen content). This finally resulted in a much higher rate of the dissolution in the melt. The starting materials were weighted in a desired ratio, mixed and pressed into pellets of 20 mm in length and 8 mm in diameter under a protective Ar atmosphere to prevent oxidation and contamination. Melting and crystallization experiments were conducted with an experimental setup, which enables simultaneous processing of Mg-Cu-B fluxes with different composition. Bars of various compositions were inserted into a graphite crucible and separated by graphite disks from each other (see Fig. 1). Before melting, the graphite crucible was inserted into stainless steel containment with 1 mm wall thickness, which was evacuated, backfilled with He and sealed to prevent Mg evaporation and maintain Mg vapour pressure at equilibrium conditions. The experimental arrangement was close to that used for $Mg_2Si$ single crystal growth reported by LaBotz and Mason [20] (see also [21]). The closed containment was installed into a tube furnace heated with 250 K/h and homogenized at a constant temperature for 20 to 50 hours. Annealing temperatures from $T_0$ = 900°C to 1350°C were selected for various runs. For crystallization the samples were slowly cooled down to room temperature with 5 K/h to 200 K/h. In order to avoid reactions between graphite and stainless steel at temperatures higher than 1100°C a Ta-foil was placed between the two crucible materials. The absence of any chemical reaction between the melted Mg-Cu-B flux and the graphite crucible was confirmed afterwards.

*2.2 Sample characterization*

The frozen samples were cut along the cylinder axis. The longitudinal sections were grinded and polished. A longitudinal section of the graphite crucible containing crystallized Mg-Cu-B samples of various compositions is shown in Fig. 1. The microstructure was revealed by optical metallography and scanning electron microscopy (SEM). Compositions of selected phases were determined by electron probe microanalysis (EPMA) applying the WDX mode. Specimens of $1 \times 1 \times 1$ $mm^3$ size containing magnesium boride crystals of different composition and morphology were cut from the frozen flux. Superconducting properties of these specimens were determined from AC-susceptibility measurements at 133 Hz and an AC-field amplitude of 1 Oe in the temperature range 5 to 55 K using both a Lakeshore model 7225 AC susceptometer and a susceptometer from the Institute of Solid State Physics RAS, Chernogolovka.

## 3. Results and discussions

### *3.1. Mg-Cu-B phase diagram*

The equilibrium phase diagram of the Mg-B binary system has not been reliably determined. The phase relations strongly depend on the Mg vapour pressure [22, 23, 24]. In particular, the $MgB_2$ phase decomposes at about 1090°C under 0.1 MPa of Mg vapour pressure. Only in the low temperature range from 650°C to 1090°C the $MgB_2$ phase may coexist with a Mg-rich melt. The Cu-B phase diagram is of simple eutectic type with no intermediate phases [22]. The maximum solubility of B in the solid Cu is about 0.3 at.%. The phase diagram of the Cu-Mg system is well established [22]. There are three eutectics and two congruently melting compounds $Mg_2Cu$ and $MgCu_2$. The solubility of Cu in Mg is restricted to 0.013 at.% Cu. The stoichiometric compound $Mg_2Cu$ melts at 568 °C. The nonstoichiometric compound $MgCu_2$ melts congruently at 797 °C and has a maximum homogeneity range of about 64.7 to 69 at.% Cu. The maximum solid solubility of Mg in Cu is 6.93 at.% Mg. No ternary Mg-Cu-B phase diagram is available so far.

### *3.2. Thermodynamic calculations*

In order to get a first idea of the possible phase equilibria, thermodynamic calculations based on the binary subsystems and on the assumption that no ternary compounds exist were carried out using the multicomponent phase diagram calculation software PANDAT [25]. The thermodynamic descriptions of the binary systems Mg-B and Cu-Mg were taken from the literature [24, 26]. The parameters for the B-Cu system were obtained empirically in order to reproduce the literature data on invariant equilibria and terminal solubilities by calculations in this work.

In the binary Mg-B system the calculations showed that the coexistence range of liquid and $MgB_2$ is extended towards higher temperature with rising pressure and at pressures exceeding 8 MPa the $MgB_2$ phase may even form directly from a Mg-rich melt via a peritectic reaction.

The first comparison of the calculated ternary phase diagram with the experimental results showed, that it is necessary to adjust the Gibbs energy description of $MgB_2$ phase as well as to take ternary interactions in the liquid phase into account. Owing to insufficient and sometimes contradictory thermodynamic data for binary Mg-B compounds [24], a more positive value (-60 kJ/mol) was proposed for the enthalpy of formation of $MgB_2$. This only results in a depression of the decomposition temperature of this compound, which has never been determined experimentally, by about 200 K. At the same time, it is in agreement with the location of fields of primary crystallization determined in the present work (see below). To estimate the behaviour of the ternary Mg-Cu-B melt calculations were performed for two limiting cases: weak interaction (Fig. 2a) and very strong interaction (Fig. 2b). Due to the limited amount of experimental data obtained in this work, the optimization of the corresponding ternary interaction parameter was not possible. Under the experimental conditions applied the solidification of the flux proceeds at elevated pressure within the sealed container. However, the Mg vapour pressure within the container is not a fixed quantity, which can be controlled properly, but it depends on the operating temperature (P ~ 0.2 MPa for 1000°C). Therefore, the projections of liquidus surface were calculated at infinitely high pressure, i.e., by excluding the gas phase from consideration. However, some general tendencies, which can be helpful for the crystal growth experiments, may be inferred. The calculations show that the crystallization range of $MgB_2$ (and of the other binary phases $MgB_4$ and $MgB_7$) has a great extent into the ternary Mg-Cu-B system. The temperatures of $MgB_2$ formation are gradually reduced with increasing Cu-fraction (cf. Fig. 2a). In the case of very strong interaction the ternary system exhibits miscibility gaps near the Mg-, Cu- and B-corners, respectively. In Fig. 2b a projection of a liquidus surface is shown, which displays both the extent of the crystallization range of $MgB_2$ and the mis-cibility gaps of the melt. The various Mg-Cu-B flux compositions treated in our experiments are indicated by various dots in Fig. 2.

### *3.3. Microstructure of as-solidified samples*

Depending on melt composition and initial temperature of the melt $T_0$, as-solidified samples contain the following phases: Mg and Cu metals, the intermetallic compounds $MgCu_2$ and $Mg_2Cu$ as well as small crystallites of Mg-borides - $MgB_2$, $MgB_4$ and $MgB_7$. Flux compositions (up to 50 at.% Cu) containing the superconducting $MgB_2$ phase in the as-solidified state are displayed in Fig. 2. No ternary Mg-Cu-B compounds were detected. The Mg-boride crystals exhibit sharply-edged geometric shapes and a dark colour and can easily be identified on the golden or silver background of the Mg-Cu flux. However, because B and Mg are light elements it is very difficult to distinguish the various Mg-borides - $MgB_2$, $MgB_4$ and $MgB_7$- from EPMA even by employing the WDX technique. In this respect the optical metallography offers great advantages. Crystals of $MgB_2$ possess strongly anisotropic characteristics of polarized light reflection and they become hardly bright if polarisation is changed. Crystals of $MgB_7$ display anisotropic properties with respect to polarized light in much less extent whereas for $MgB_4$ crystals almost no optical anisotropy was observed.

Besides the flux composition, the initial melt temperature $T_0$ before slow cooling turned out as the decisive parameter for the formation of Mg-borides. For high temperatures ($T_0$ > 1100°C) the formation of $MgB_4$ and $MgB_7$ prevails and no $MgB_2$ was formed. Typical microstructures of a $Mg_{30}Cu_{50}B_{20}$ flux cooled by 5 K/h from 1200°C containing non-superconducting $MgB_7$ and $MgB_4$ precipitates are shown in Fig. 3. The $MgB_7$ particles display a plate-like shape with typical dimensions $350 \times 350 \times 50$ µm$^3$. The $MgB_4$ particles tend to be equiaxed sharply edged grains up to 500 µm in size.

At temperatures below $T_0 \approx 1100$°C for the same $Mg_{30}Cu_{50}B_{20}$ melt composition the $MgB_2$ phase can be formed peritectically as a rim surrounding the $MgB_4$ primary phase as demonstrated in Fig. 4a. That confirms that the theoretically predicted peritectic reaction

$L + MgB_4 \rightarrow MgB_2$

at elevated pressure indeed proceeds for ternary Mg-Cu-B alloys. So far, the peritectic formation of $MgB_2$ was not observed. If the initial melt temperatures were below 1100°C, for melt compositions in the range $Mg_{80-30}Cu_{10-50}B_{10-20}$ (*cf*. Fig. 2), the $MgB_2$ phase is primarily crystallized from the Mg-Cu-B flux. As shown in Fig. 4b and 4c the $MgB_2$ crystals appear as thin plates and their sizes strongly depend on melt composition and temperature course. The maximum size of the crystal achieved is about $200 \times 200 \times 50$ µm$^3$ for the optimum process conditions ($Mg_{50}Cu_{40}B_{10}$, $T_0 \approx$ 1050°C, and cooling rate 5 K/h). The absence of any grain boundaries within the $MgB_2$ particles provide strong evidence of their single crystalline nature. Very often MgO particles are incorporated within the matrix of $MgB_2$ crystals. They probably act as seeds for the generation of new $MgB_2$ crystals and strongly reduce the maximum crystal size. So far we have not found appropriate techniques to get rid of solid Mg-oxide in the Mg-Cu-B melt.

For Mg-rich flux compositions with < 30 at.% Cu a phase separation of the melt into Mg-rich and Cu-rich regions was inferred from the microstructure. This is illustrated in Fig. 5 for a $Mg_{70}Cu_{10}B_{20}$ flux. There is an obvious tendency of forming Mg-rich regions with a high density of small $MgB_2$ particles separated by a Cu-rich flux containing larger $MgB_2$ particles with lower particle number density. Seemingly, this microstructure arises from a miscibility gap in the Mg-rich corner as predicted by the Mg-Cu-B phase diagram calculation shown in Fig. 2b. The concentration range for phase separation phenomena observed in microstructures does not quantitatively agree with the predicted miscibility gap but extends toward somewhat higher Cu contents.

In agreement with thermodynamic calculations our experimental observations suggest that the vapour pressure of Mg must have a great influence on $MgB_2$ phase formation from the melt. Any attempt of directional solidification of Mg-Cu-B melts in a finite temperature gradient (30 K/cm) failed to form $MgB_2$, instead the $MgB_4$ phase was obtained. We interpret this failure as a result of the lower temperature at cooled side of the sealed container, which may cause sublimation of the Mg vapour. This would immediately result in a reduced equilibrium Mg vapour pressure in the ampoule and finally, in the decomposition of the $MgB_2$ phase within the Mg-Cu-B melt flux. That renders the operating window of $MgB_2$ phase formation from the Mg-Cu-B melt flux extremely narrow.

*3.3 Superconducting properties*

Superconducting properties have been measured for a wide composition range of $Mg_{80-30}Cu_{10-50}B_{10-20}$ flux bars containing $MgB_2$ particles. Within the Mg-Cu-B system only the $MgB_2$ phase is known to display superconducting properties with $T_c$ > 4.2 K. That strongly simplifies the data analysis of superconductivity measurements. The superconducting behaviour detected in bulk samples originate from $MgB_2$ particles crystallized from the molten Mg-Cu-B flux. All flux compositions which exhibit no superconducting properties are indicated by open dots in Fig. 2. No traces of unreacted $MgB_2$ powder from the starting materials have been detected, which might have contributed to the superconducting behaviour. For great excess of boron (> 20 at.% B) in the flux composition $MgB_4$ particles have been detected, which have not been precipitated from the melt but are likely the result of $MgB_2$-powder decomposition processes during heating.

Typical AC susceptibility vs. temperature plots $c$(T) are shown in Fig. 6. No superconductivity beyond 4.2 K has been detected in specimens without macroscopically visible $MgB_2$ particles. The largest superconducting transition temperature $T_c$ = 39.2 K and the narrowest transition width $DT_c$ = 1.3 K (from 90% to 10% of $c$(T)) have been observed in Mg-Cu-B specimens containing $MgB_2$ particles formed by a peritectic reaction from the $MgB_4$ phase (cf. Fig. 4a). This phase formation mode was found for Cu-rich fluxes ($Mg_{30}Cu_{50}B_{20}$) and annealing temperatures $T_0 \approx$ 1100°C . The cooling rate plays a minor role in the superconducting phase formation. Flux compositions in the range of $Mg_{80-40}Cu_{10-40}B_{10-20}$ with plate-like $MgB_2$ crystals primarily formed at liquidus temperatures below the peritectic point, exhibit superconducting transitions at lower $T_c$ possibly depending on the temperature of formation of the $MgB_2$ phase. As shown in Fig. 6, typically, two subsequent superconducting transitions have been inferred from the measurements for Mg-rich flux composi-tions in the vicinity of the predicted miscibility gap (*cf.* Fig. 2b). The sample $Mg_{60}Cu_{30}B_{10}$ (1) with small B-concentration displays two transitions with low onset temperatures $T_{c1} \approx 22$ K and $T_{c2}$ = 7.2. For various specimens along an isoplethal section at 20 at.% B from 70 to 50 at.% Mg we derived $T_{c1} \approx 39$ K and a very distinct second transition at $T_{c2}$ , which slopes down from 31 to 27 K with increasing Cu-content of the flux from 10 to 30 at.%. But for $Mg_{40}Cu_{40}B_{20}$ the second transition practically disappears, which gives rise to a smooth transition similar to that of crystals prepared by vapour deposition [18].

The possible effect of Cu solubility on superconducting properties of $MgB_2$ [14,15] was revealed by EPMA analysis. No trace of Cu has been detected in $MgB_2$ particles. Therefore we can rule out any Cu-solubility in $MgB_2$ within the accuracy limits of the WDX analysis (< 1 at.%). The high onset temperature $T_{c1} \approx 39$ K itself excludes sizeable Cu solubility because of the reported depression of $T_c$ by 0.2 K/mol.% Cu [14]. The existence of a finite homogeneity region of $MgB_2$ compound is assumed to explain the variation of superconducting transition temperatures $T_c$. This is already suggested by the large transition width of the order of $DT_c$ ~ 10 K of crystals prepared by vapour deposition [18]. The highest $T_c$ was achieved for the peritectically formed $MgB_2$ that is assumed to possess a composition close to stoichiometry. The smaller temperature of formation of the $MgB_2$ crystals the larger is the deviation from stoichiometry, accordingly resulting in lower $T_c$. The two very distinct superconducting transitions in phase separated samples hint to growth processes proceeding at different temperatures within the Mg- and the Cu-enriched flux regions, respectively, and accordingly to different compositions of the individual $MgB_2$ crystals. A particularly, low $T_c$ is achieved in the $Mg_{60}Cu_{30}B_{10}$ sample with low B-concentration which may lead to B-depletion in $MgB_2$ crystals. Our suggestion also coincides with conclusions from superconducting measurements of ceramic $MgB_2$ samples [16] and on Mg-enriched $MgB_2$ films with very low $T_c$ =12 K [17].

## 4. Conclusions

A new method is presented for growth of $MgB_2$ crystals from Mg-Cu-B melt flux. Large solubility of > 20 at.% B in the Mg-Cu melt at temperatures > 1000°C, depending on melt composition, have been found. First estimates of ternary Mg-Cu-B phase diagram by thermodynamic calculations show the beneficial effect of Cu and elevated pressure on $MgB_2$ phase formation. Unlike the binary Mg-B system, $MgB_2$ phase formation from the melt is enabled by reducing the liquidus temperature. The theoretically predicted phase separation phenomena in the Mg-rich region has been observed experimentally. The size of the $MgB_2$ crystals strongly depends on melt composition and initial melt temperature $T_0$. Although not quantified the elevated pressure within the sealed container seems to be indispensable for $MgB_2$ phase formation. The maximum size of single crystals of 200 x 200 x 50 $\mu m^3$ for a flux composition $Mg_{50}Cu_{40}B_{10}$ and $T_0$ = 1050°C was achieved. MgO particles contained in the melt seem to be one of the limiting factors for crystallite size since they may act as seeds for $MgB_2$ crystallization.

Two different crystallization modes of $MgB_2$, a peritectic reaction with $MgB_4$ and primary growth of $MgB_2$ particles from the melt, have been identified depending on initial melt temperature and flux composition. From the analysis of Mg-Cu phases around $MgB_2$ particles and the starting melt composition a rough estimate of $MgB_2$ phase formation temperatures (between 780°C and 1000°C) can be inferred.

So far the superconducting properties of $MgB_2$ crystals can only be deduced from AC-susceptibility measurements of Mg-Cu-B flux bars. The highest $T_c$ = 39.2 K and a narrow transition width have been measured in the specimen containing peritectically formed $MgB_2$ particles assumed to be close to stoichiometric composition. The lower the liquidus temperature of primarily formed plate-like $MgB_2$ single crystals the larger is the deviation from stoichiometry resulting in lower $T_c$ and extended transition widths. This behaviour suggests a finite homogeneity range of the $MgB_2$ compound. Samples which display a liquid immiscibility typically exhibit two supercon-ducting transitions. A sizeable effect of Cu solubility on superconducting properties of $MgB_2$ can be ruled out. Measurements on isolated $MgB_2$ single crystals prepared from the Mg-Cu-B flux are under way.

**Acknowledgements**

The authors would like to thank M. Frömmel and Müller-Litvanyi for technical support and S. Pichl for EPMA measurements. The work was performed partially with financial support of the Deutsche Forschungsgemeinschaft (DFG) within SFB 463 „Rare-Earth Intermetallics: Structure, Magnetism and Transport“.

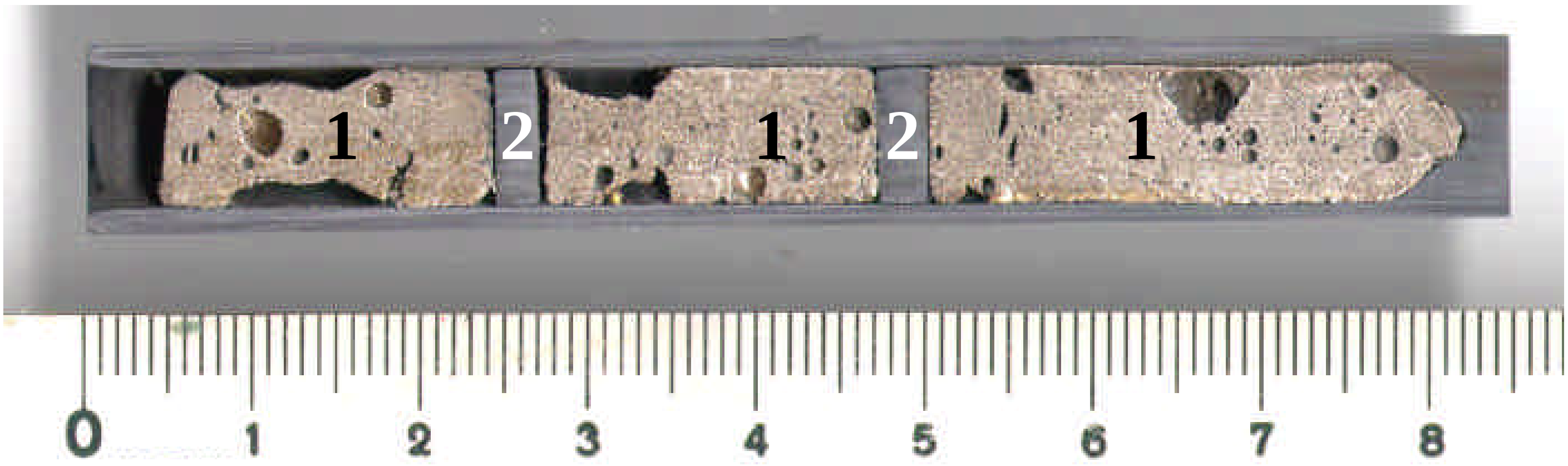

Fig. 1.  Longitudinal section of a graphite crucible of 8.5 mm inner diameter with 3 bars of solidified Mg-Cu-B flux (1) containing $MgB_2$ crystals. The Mg-Cu-B bars are separated by graphite disks (2).

a)

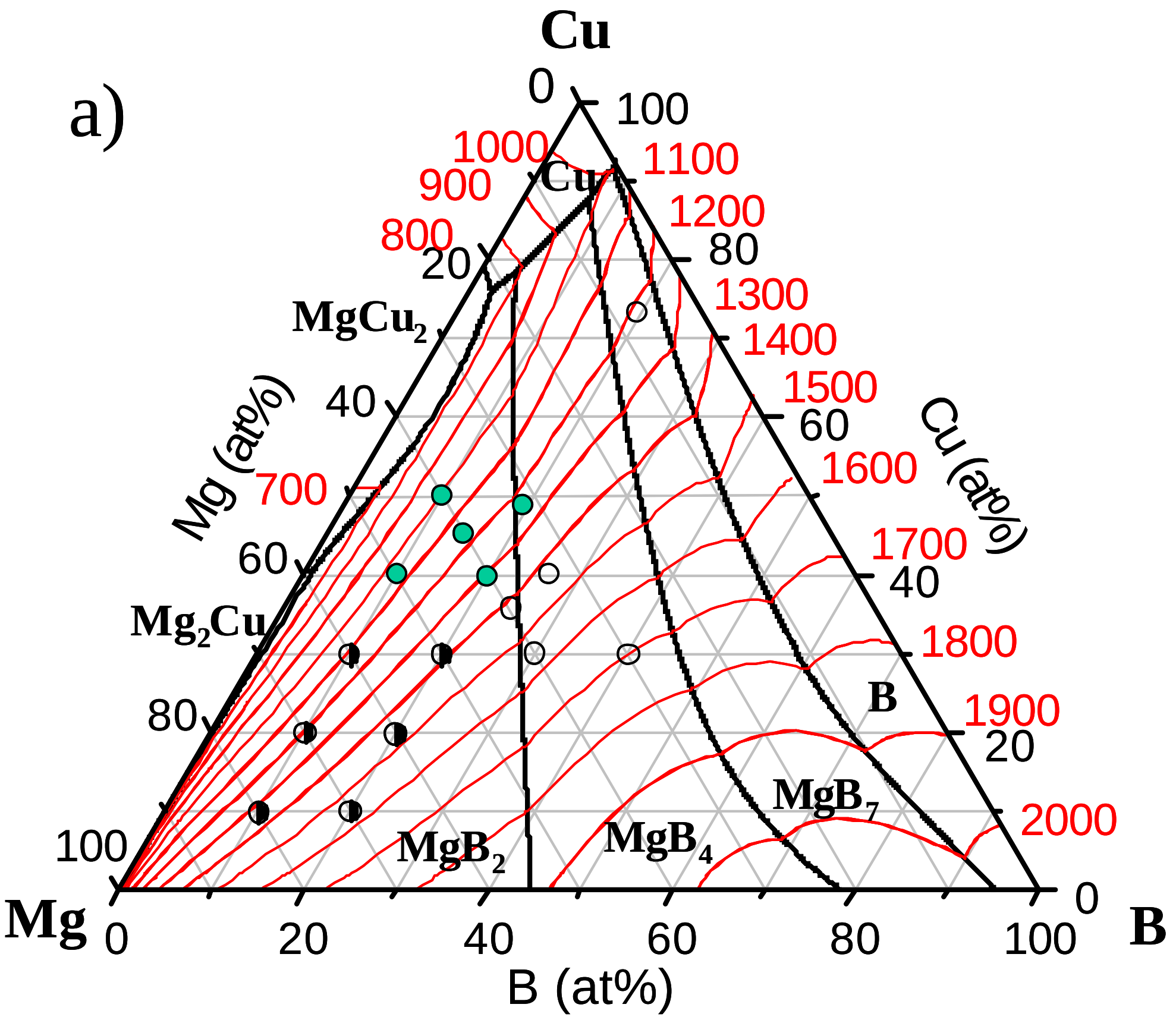

Fig. 2a

Fig. 2. Calculated projections of liquidus surface of the ternary Mg-Cu-B phase diagram at infinitely high pressure based on the thermodynamics of binary subsystems and two different assumptions: (a) phase diagram showing the primary solidification ranges of the respective phases and the liquidus isotherms (in °C) for weak ternary interaction in the liquid; (b) phase diagram showing the primary solidification ranges of the respective phases and the existence of liquid miscibility gaps in the Mg-, Cu- and B-rich corner for very strong ternary interaction in the liquid. The dots mark the Mg-Cu-B flux compositions studied experimentally. Full dots: fluxes containing $MgB_2$ crystals; half-filled dots: samples with liquid phase separation

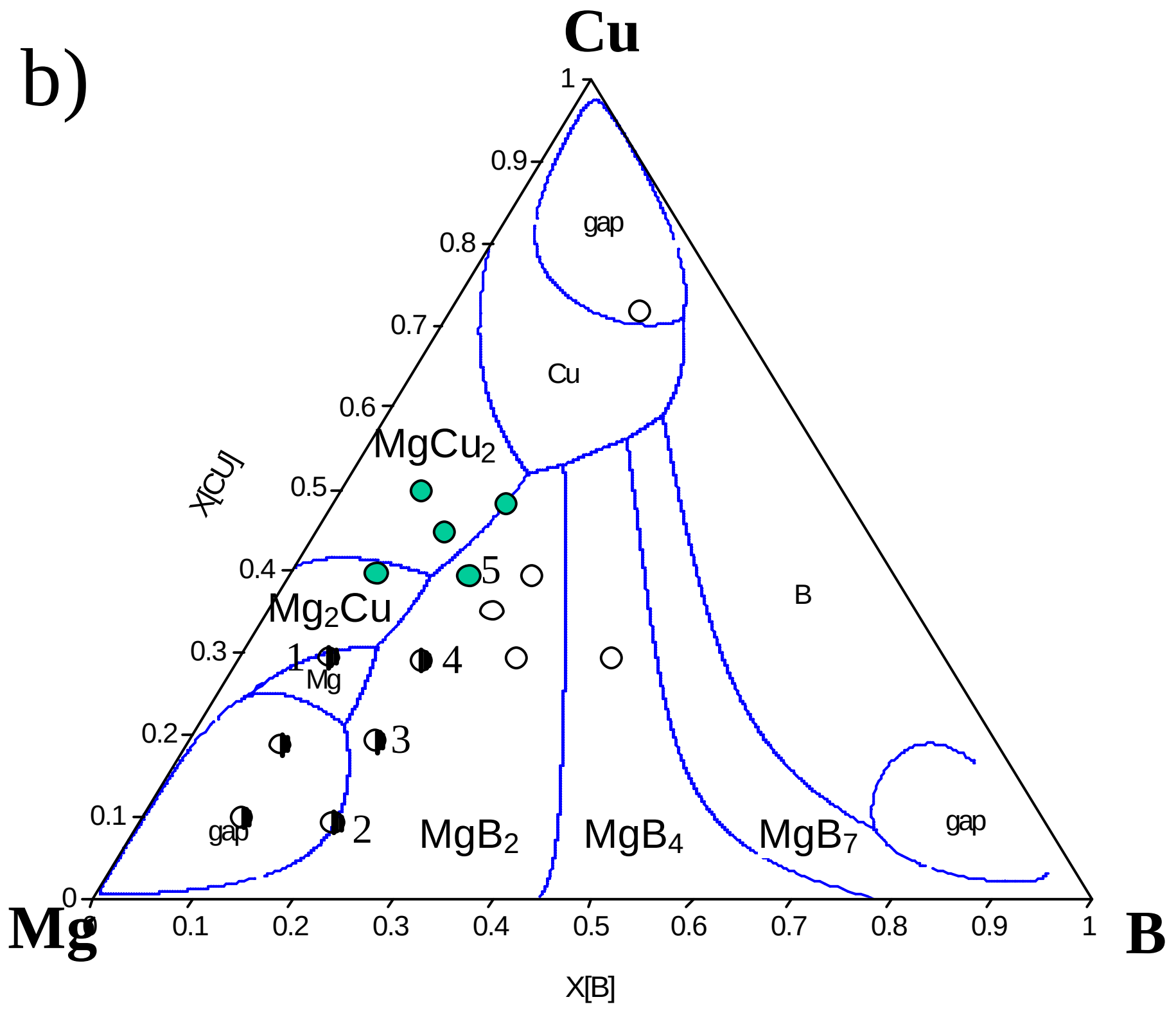

Fig. 2b

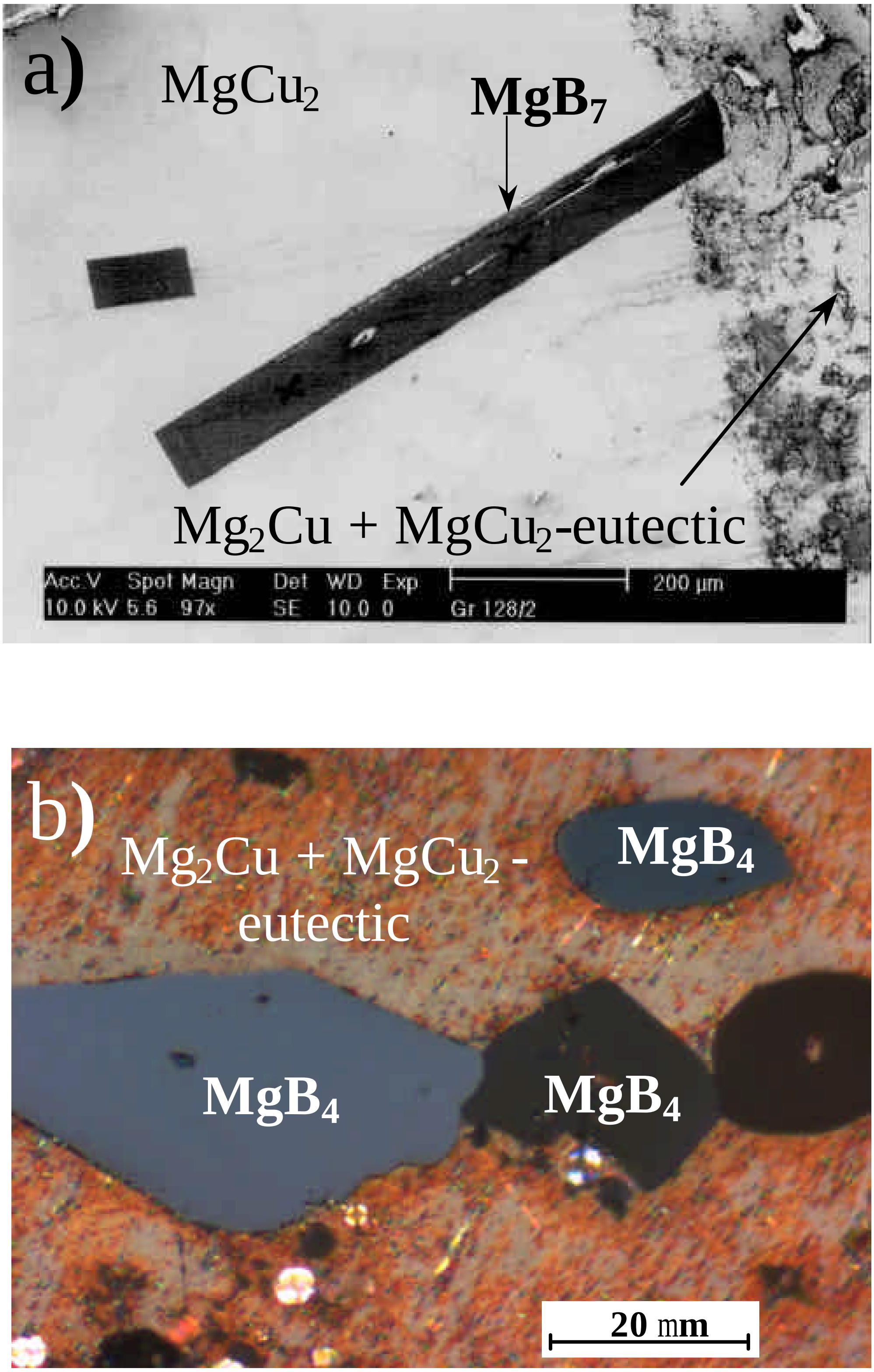

Fig. 3. Microstructure of a $Mg_{30}Cu_{50}B_{20}$ flux cooled from high temperatures $T_0$ = 1200°C with 5 K/h, which contains non-superconducting $MgB_7$ (a) and $MgB_4$ (b) precipitates, respectively.

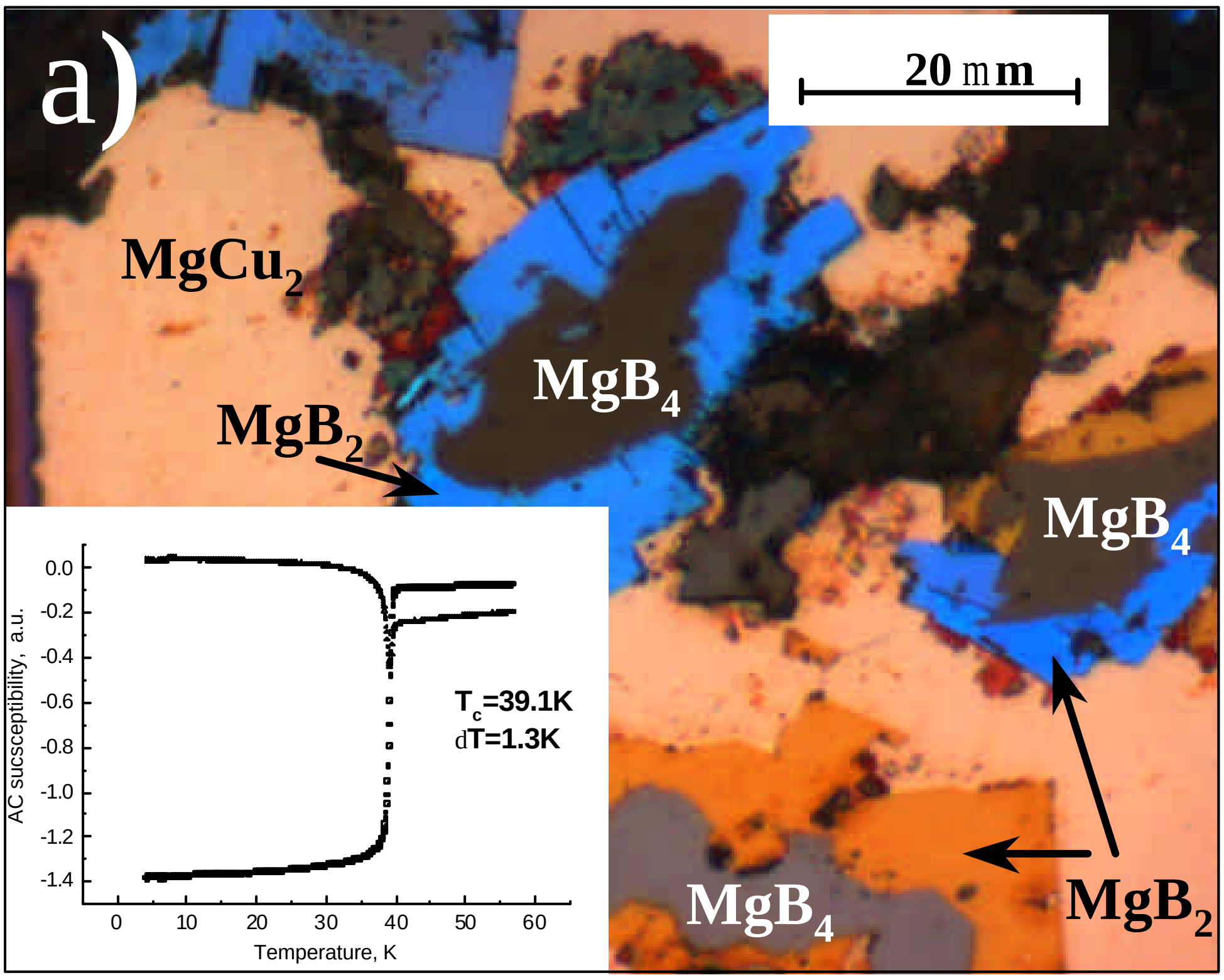

Fig. 4a

Fig. 4. $MgB_2$ crystallized from melt fluxes of different composition at different growth conditions. (a) The peretectically formed $MgB_2$ phase (bright) around a $MgB_4$ primary precipitate (dark) for $Mg_{30}Cu_{50}B_{20}$ and $T_0$ = 1100°C. Inset: measured AC susceptibility *vs*. temperature $\chi$(T) of the sample with an onset temperature of superconductivity of $T_{c0}$ = 39.2 K . (b) Large plate-like primary $MgB_2$ crystals for $Mg_{35}Cu_{40}B_{25}$ and $T_0$ = 1050°C. (c) Small plate-like primary $MgB_2$ crystals for $Mg_{50}Cu_{30}B_{20}$ and $T_0$ = 1050°C.

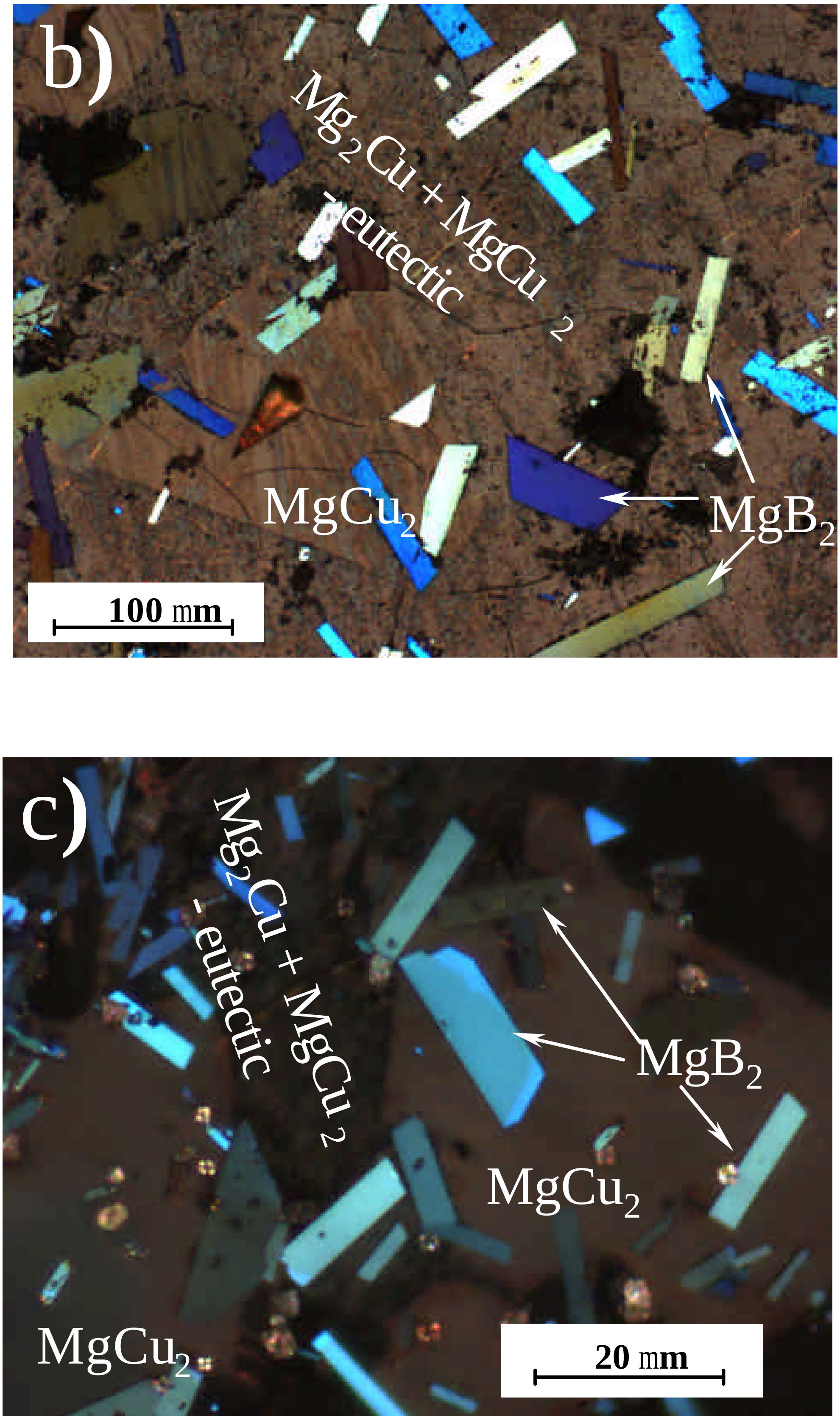

Fig. 4b, c

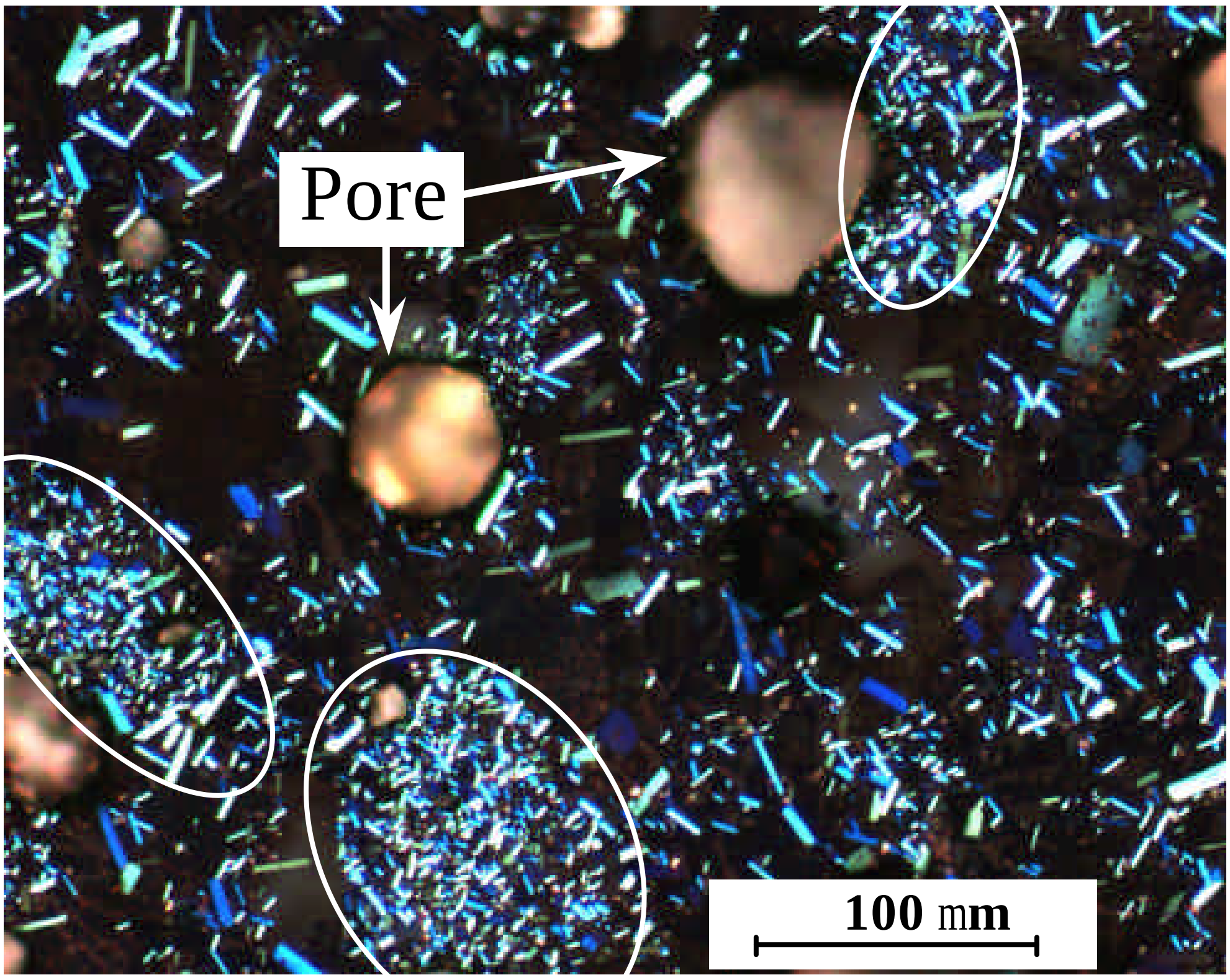

Fig. 5. Sample of composition $Mg_{70}Cu_{10}B_{20}$ ($T_0$ = 1050°C) showing separation into Mg-rich regions (encircled) displaying high particle density of small $MgB_2$ crystals and Cu-enriched regions with larger $MgB_2$ crystals but low particle density.

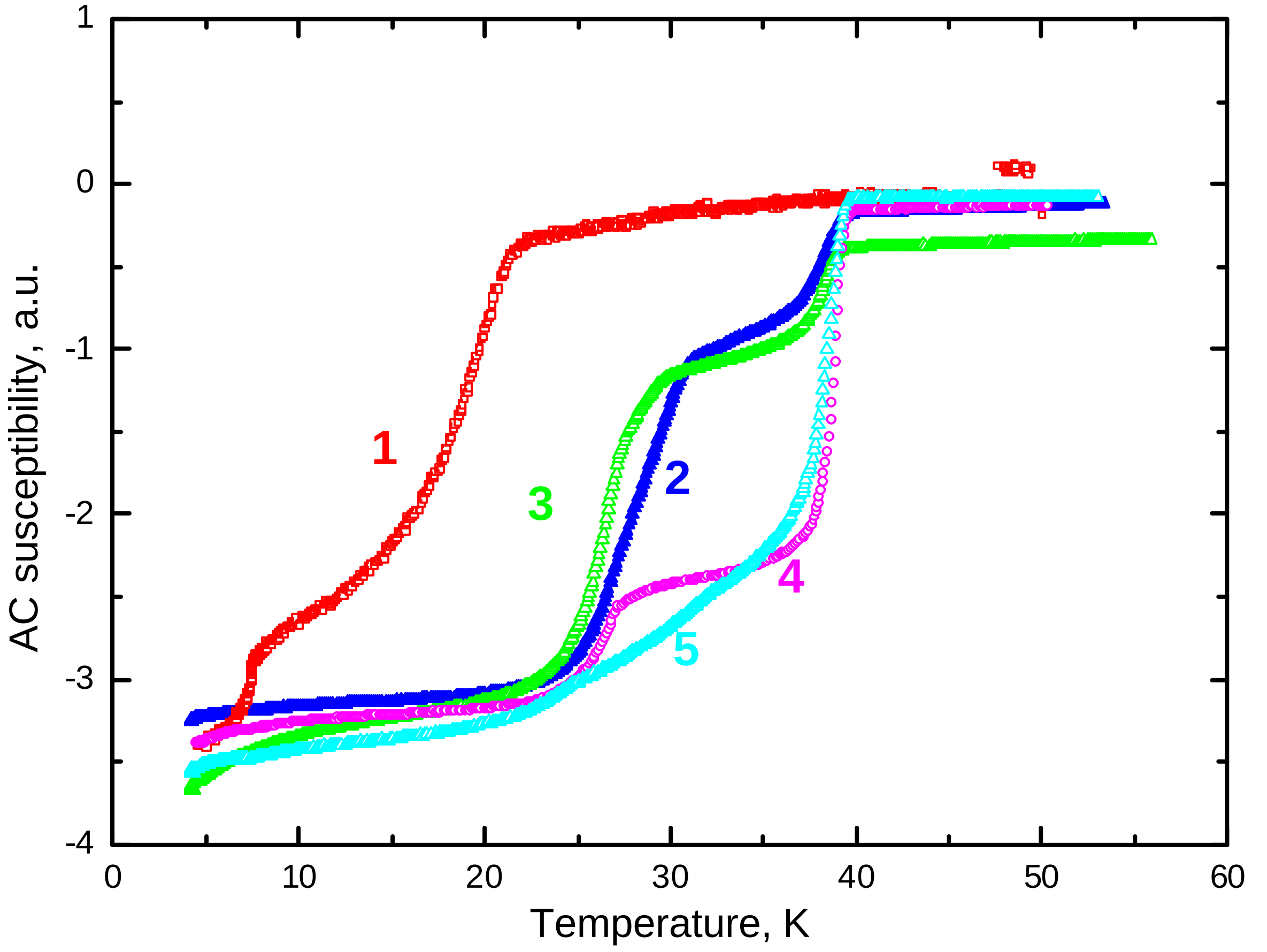

Fig. 6. AC susceptibility vs. temperature χ(T) showing superconducting transitions in samples with phase separation $Mg_{60}Cu_{30}B_{10}$ (1), $Mg_{70}Cu_{10}B_{20}$ (2), $Mg_{60}Cu_{20}B_{20}$ (3), $Mg_{50}Cu_{30}B_{20}$ (4), and with no phase separation $Mg_{40}Cu_{40}B_{20}$ (5) prepared for initial melt temperatures $T_0$ = 1050°C.